\documentclass[aps, pra, showpacs, amsmath, twocolumn, groupedaddress]{revtex4}
\usepackage{graphicx}
\usepackage{calc}
\usepackage{amsmath}
\graphicspath{{figdistilled/}{./}}

\RequirePackage{amsmath}

\renewcommand{\vec}[1]{\ensuremath{\boldsymbol{#1}}}
\newcommand{\vhat}[1]{\ensuremath{\hat{\vec{#1}}}}
\newcommand{\unit}[1]{\ensuremath{\,\mathrm{#1}}}

\newcommand{\ham}{\ensuremath{H}}

\newcommand{\hilb}{\ensuremath{\mathcal{H}}}
\newcommand{\fid}{\ensuremath{\mathcal{F}}}

\newcommand{\ke}{\ensuremath{\ket{\text{e}}}}
\newcommand{\be}{\ensuremath{\bra{\text{e}}}}
\newcommand{\kl}{\ensuremath{\ket{1}}}
\newcommand{\ko}{\ensuremath{\ket{0}}}
\newcommand{\gt}{\ensuremath{g_\text{t}}}
\newcommand{\kaux}{\ensuremath{\ket{\text{aux}}}}

\newcommand{\pulse}{P}

\newcommand{\bra}[1]{\left\langle{#1}\right\rvert}
\newcommand{\ket}[1]{\left\lvert{#1}\right\rangle}
\newcommand{\expval}[1]{\left\langle{#1}\right\rangle}
\newcommand{\Braket}[1]{\expval{#1}}
\newcommand{\pp}[3]{^{#3}\pulse^{#1}_{#2}}

\begin{document}

\title{
Robust quantum gates and a bus architecture for quantum computing with
rare-earth-ion doped crystals.
}

\author{Janus Wesenberg}
\email[]{jaw@phys.au.dk}
\author{Klaus M{\o}lmer}
\affiliation{
  QUANTOP,
  Danish Research Foundation Center for Quantum Optics,
  Department of Physics and Astronomy, 
  University of Aarhus,
  DK-8000 {\AA}rhus C, 
  Denmark}

\date{\today}

\begin{abstract}
  We present a composite pulse controlled phase gate which together
  with a bus architecture improves the feasibility of a recent quantum
  computing proposal based on rare-earth-ion doped crystals.
  Our proposed gate operation is tolerant to variations between ions of
  coupling strengths, pulse lengths, and frequency shifts, and it achieves
  worst case fidelities above $0.999$ with relative variations in
  coupling strength as high as $10\%$ and frequency shifts up to several
  percent of the resonant Rabi frequency of the laser used to
  implement the gate.
  We outline an experiment to demonstrate the creation and detection
  of maximally entangled states in the system.
\end{abstract}



\maketitle
\section*{\label{sec:introduction}Introduction}
The Rare Earth Quantum Computer (REQC) recently proposed by Kr{\"o}ll
and coworkers is a solid state ensemble quantum computer based on
laser-addressed rare earth ions embedded in cryogenically cooled
inorganic crystals
\cite{ohlssona02:quant_comput_hardw_based_rare,nilsson02:initial_exper_concer_quant_infor}.
Internally, the system consists of a macroscopic number of
independent \emph{instances} of the same quantum computer.
The qubits, represented by ground state hyperfine levels in the rare
earth ions, are divided into \emph{channels} according to the
inhomogeneous frequency shifts of an optical transition in the ions.
Every instance of the quantum computer has one representative from
each active channel, and the instances are operated in parallel by
addressing the channels with resonant optical radiation.

The inhomogeneous shift of the ions in a given channel will not be
quite identical, and there will in general be fluctuations in their
coupling to an applied coherent field.
Also, the static dipole coupling used to mediate multi qubit gate
operations will differ between instances.
As a consequence, we must use gate operations that are stable with
respect to variations in these parameters which, as we will
demonstrate in this paper, may be achieved by using composite pulses
and phase compensating operations.

One of the exciting features of REQC is that the size and coupling
topology of the quantum computer is not defined by the crystal, but
rather chosen in an initialization stage at each start-up of the
system.
The choice of architecture determines the number of instances
available in a given crystal, and thus ultimately the scaling
properties of the system.
We propose to choose a bus-based architecture, as this will both allow
for a higher number of instances and simplify the gate implementation
as compared to the originally proposed architecture.

We conclude the paper with a proposal for an experimental
demonstration of creation and detection of maximally entangled states
in REQC systems.

\section{Quantum computing with rare earth ions}
\label{sec:reqc-proposal}
Rare earth ions embedded in cryogenic crystals have a number of
features making them suitable for quantum information processing
\cite{longdell02:exper_demon_quant_state_tomog}:
\begin{itemize}
\item Ground state hyperfine levels with a lifetime of hours and
  decoherence times up to several $\unit{ms}$.  We will use three such
  states, labeled $\ko$, $\kl$, and $\kaux$, to implement quantum
  registers and for parking unwanted ions.
\item Optical transitions with homogeneous line widths on the order
  of $\unit{kHz}$ are inhomogeneously broadened to several $\unit{GHz}$,
  allowing us to address a large number of independent channels.
\item The crystal-embedded ions have large static dipole moments with
  interaction energies up to several $\unit{GHz}$.
  This interaction is ideal for implementing gate operations of the
  dipole blockade type.
\end{itemize}

In the remainder of this section we will briefly introduce the basic
ideas of REQC, as originally described in
Ref.~\cite{ohlssona02:quant_comput_hardw_based_rare}.

\subsection{\label{sec:dynam-arch-select}Dynamical architecture selection}
The architecture of the REQC system is selected at start-up by an
initialization procedure.  The desired end-point of this process is a
large number of independent instances of the chosen quantum computer,
each instance being a group of ions with one representative from each
active channel and couplings between the ions as required by the
chosen architecture.

The initialization proceeds in two steps: channel preparation and
identification of quantum computer instances.
In both of these steps unwanted ions are deactivated by transferring
them to off-resonant, metastable states.

\paragraph{Channel preparation.}
A channel refers to a large number of ions distributed throughout the
crystal, all having the same inhomogeneous shift and coupling strength
within the inhomogeneously broadened optical transition used to access
the ions.
The channel preparation aims to deactivate all dopant ions close to
resonance with a given channel and to transfer all members of the
channel itself to their $\ket{0}$ state.

This can be achieved by means of spectral hole burning techniques, and
widths of the final channel structure as low as $50 \unit{kHz}$ have
been obtained experimentally for materials similar to those considered
for use in REQC \cite{longdell02:exper_demon_quant_state_tomog}.

\paragraph{Instance identification.}
After a successful initialization, each ion will only be interacting
with ions from other channels, allowing us to ignore ``excitation
hopping'' transitions \cite{lukin00:quant_entan_optic_contr_atom},
as these will not be energy conserving.
As a consequence we can model the dipole coupling as simple couplings
between the excited states:
\begin{equation}
  \label{eq:couplingb}
  V_\text{dipole}=\frac{1}{2}\sum_{\mu \neq \nu} g_{\mu \nu}
  \left(\ke\be\right)_\mu \otimes
  \left(\ke\be\right)_\nu,
\end{equation}
where the sum is over all pairs of ions.
To be precise about the objectives of the instance identification
process, we will consider ions $\mu$ and $\nu$ to be coupled if
$g_{\mu \nu}$ exceeds a threshold $g_t$ determined by the chosen
implementation of the gate operation.

The goal of the instance identification procedure is to transfer ions,
which are in an active channel but not members of a valid instance,
to their auxiliary state $\kaux$.
One way to achieve this is to go through the following procedure 
for each pair, $(i,j)$, of channels required to be coupled:

By applying a $\pi$-pulse to ions in channel $i$ we transfer the
$\ko$ population to the $\ke$ state, thus shifting the excited state
energy of all ions coupled to a channel $i$ ion.
By means of a frequency sweep or a comb of $\pi$ rotations, all
channel $j$ ions which are shifted less than $\gt$ are now transfered
to their excited state $\ke$, after which the channel $i$ ions are
returned to $\ko$.
We now wait for the excited channel $j$ ions to decay, which will
transfer part of the ions to the inactive $\kaux$ state.

By repeated application of this pulse sequence, we can deactivate an
arbitrarily high fraction of the channel $j$ ions which are not
coupled to a channel $i$ ion.
After this has been achieved, we repeat the process with the roles of
channel $i$ and $j$ interchanged, and afterwards proceed to apply the
same procedure to all other edges of the coupling graph to finally arrive at
the desired initialized REQC system.

\subsection{\label{sec:gate-operation}Gate operation}
In general, the coupling strengths, $g_{\mu \nu}$, will differ
between instances, requiring us to use gate operations that do not
depend on the precise magnitude of the coupling strength. 

One gate operation with this quality is the controlled phase shift
based on the dipole blockade effect
\cite{jaksch00:fast_quant_gates_neutr_atoms}.
Assuming all ions not participating in the operation to be in their
qubit states, $\ko$ and $\kl$, and thus decoupled from the operation,
we can implement a controlled phase shift in its simplest form by the
following pulse sequence:
\begin{equation}
  \label{eq:phasegate} 
  \pulse^{(i)}_{0e}(\pi,\pi) \, 
  \pulse^{(j)}_{1e}(0,2\pi) \, 
  \pulse^{(i)}_{0e}(0,\pi),  
\end{equation}
with $\pulse^{(i)}_{ab}(\phi,\theta)$ representing the effect of a
resonant pulse of area $\theta$ and phase $\phi$ applied on the
$\ket{a}-\ket{b}$ transition of ions in channel $i$.

For two coupled ions $\mu$ and $\nu$, residing in channels $i$ and $j$
respectively, the effect of performing the pulse sequence
\eqref{eq:phasegate} would be the following:
If ion $\mu$ is initially in the $\ko_\mu$ state, the
$\pulse^{(i)}_{0e}(0,\pi)$ pulse transfers the ion to the excited
state $\ke_\mu$ and thus shifts the $\pulse^{(j)}_{1e}(0,2 \pi)$ pulse
out of resonance, causing the system to return to the initial state
after the last $\pi$ pulse.
If, on the other hand, ion $\mu$ is initially in the $\kl_\mu$ state,
it is not transferred to $\ke_\mu$ and the $2 \pi$ pulse is resonant
and causes a $\pi$ phase shift on the $\kl_\nu$ state.
The effect of the full gate operation on the qubit space is
consequently a $\pi$ phase shift on the $\ket{11}$ state:
$U_\text{CPS}=1-2\ket{11}\bra{11}$.

\section{High fidelity gate operations}
\label{sec:gate-operations}
Gate operations for the REQC system face a number of challenges due to
the fact that they operate simultaneously on a number of not quite
identical instances of a quantum computer:
Due to the finite channel width, ion $\mu$ will in general be detuned
by a small amount $\delta^{(\mu)}$ from the central channel frequency.
Furthermore, the experienced Rabi frequency, $\Omega_0^{(\mu)}$, will
differ slightly from the average Rabi frequency $\Omega_0$ due to
laser field inhomogeneities and local variations in dipole moments.

In this section we will show that by taking advantage of the fact that
$\delta^{(\mu)}$ and $\Omega_0^{(\mu)}/\Omega_0$ are constant in time
for each ion, we can design pulse sequences that perform almost the
same operation on each instance.

\subsection{Composite rotations}
\label{sec:composite-rotations}
The pulse $\pp{}{ie}{}(\phi,\theta)$ is driven by a Hamiltonian
$\tilde{\ham}_1=\frac{1}{2}\vec{\Omega}\cdot \vec{\sigma}^{(ie)}$ with
$\vec{\sigma}^{(ie)}$ signifying the Pauli-matrices in the
$\{\ket{i},\ke\}$ basis and $\vec{\Omega} = \Omega_0 \vhat{n}_\phi$,
where $\vhat{n}_\phi$ is a unit vector in the $x-y$ plane with
azimuthal angle $\phi$.

To apply the pulse $\pp{}{ie}{}(\phi,\theta)$ we engage the field for
a period $\theta/\Omega_0$, so that an ideal reference ion, $\xi$,
with $\delta^{(\xi)}=0$ and $\Omega_0^{(\xi)}=\Omega_0$ will be
rotated by an angle $\theta$ around $\vhat{n}_\phi$ as desired.
In general, however, the ions will react differently to the pulse due
to their different detunings and coupling strengths. 

The problem of taking all the ions through the same evolution when
they react differently to the pulses has been studied in great detail
in the magnetic resonance community \cite{levitt86:compos_pulses}.
Inspired by the discussion in
Ref.~\cite{cummins02:tackl_system_error_quant_logic} we have used the
\emph{BB1} pulse sequence to replace a single pulse $\pulse(0,
\theta)$ with the following sequence of pulses:
\begin{multline}
  \label{eq:ubb1}
  \pulse_\text{BB1}(0,\theta)=\\
  \pulse(0,\theta/2) \, 
  \pulse(\phi_c,\pi) \,
  \pulse(3 \phi_c,2 \pi)\, 
  \pulse(\phi_c,\pi) \,
  \pulse(0,\theta/2).
\end{multline}
For our reference ion, $\xi$, the unitary evolution
${^\xi\pulse_\text{BB1}}(\phi,\theta)$ caused by the
$\pulse_\text{BB1}(\phi,\theta)$ composite pulse is seen to be exactly
identical to the evolution caused by $\pulse(\phi,\theta)$.
The use of five pulses for this simple task is justified, however, if
we instead consider the evolution ${^\mu\pulse_\text{BB1}}$ of a
general ion subject to the Hamiltonian 
\begin{equation}
  \label{eq:errorham1}
  \ham_1^{(\mu)}=-\delta^{(\mu)}\,\ke\be
  +\frac{1}{2}\Omega_0^{(\mu)} \vhat{n}_\phi \cdot \vec{\sigma}^{(ie)}.
\end{equation}
In this case we find that with the optimal value,
$\phi_c=\pm \cos^{-1}\left(-\theta/4 \pi\right)$,
${^\mu\pulse_\text{BB1}}(\phi,\theta)$ is almost constant over a large
range of values of $\delta^{(\mu)}/\Omega_0$ and
$\Omega^{(\mu)}/\Omega_0$, while $\pp{}{}{\mu}(\phi,\theta)$ changes
quite rapidly.

%
%

\subsection{Robust gate operation}
\label{sec:full-fledged-gate}
For the two-level Rabi problem there is a global phase factor
depending on the detuning which plays no observable role. In our
three-level system, however, this phase will lead to a dephasing
between the qubit level $\ket{i}$ coupled to $\ke$ and the other qubit
level.
To compensate this, we must symmetrize the desired pulse sequence in a
suitable way, to allow both levels to pick up the same, unknown, phase
contributions.

In the case of the controlled phase shift \eqref{eq:phasegate}, we
have arrived at the following symmetrized version:
\begin{multline}
  \label{eq:cphase}
  \pulse^{(i,j)}_\text{CPS}=
  \pulse^{(i)}_{1e}(\pi,\pi) \\ 
  \pulse^{(j)}_{0e}(\pi,\pi) \, 
  \pulse^{(j)}_{0e}(0,\pi) \, 
  \pulse^{(j)}_{1e}(\pi,\pi) \, 
  \pulse^{(j)}_{1e}(0,\pi) \\ 
  \pulse^{(i)}_{1e}(0,\pi) \,
  \pulse^{(i)}_{0e}(\pi,\pi) \\ 
  \pulse^{(j)}_{0e}(\pi,\pi) \, 
  \pulse^{(j)}_{0e}(0,\pi) \, 
  \pulse^{(j)}_{1e}(0,\pi) \, 
  \pulse^{(j)}_{1e}(0,\pi) \\ 
  \pulse^{(i)}_{0e}(0,\pi).
\end{multline}
For the reference ion $\xi$ the $\pulse^{(i,j)}_\text{CPS}$ pulse
sequence is seen to be equivalent to $\pulse^{(i)}_{0e}(\pi,\pi) \,
\pulse^{(j)}_{1e}(0,2\pi) \, \pulse^{(i)}_{0e}(0,\pi)$, which is
exactly the basic controlled phase shift operation
\eqref{eq:phasegate}, but we expect it to perform better for a general
ion.

\begin{figure*}[htbp]
  \begin{tabular}{ll}
    (a)&(b)\\
    \includegraphics{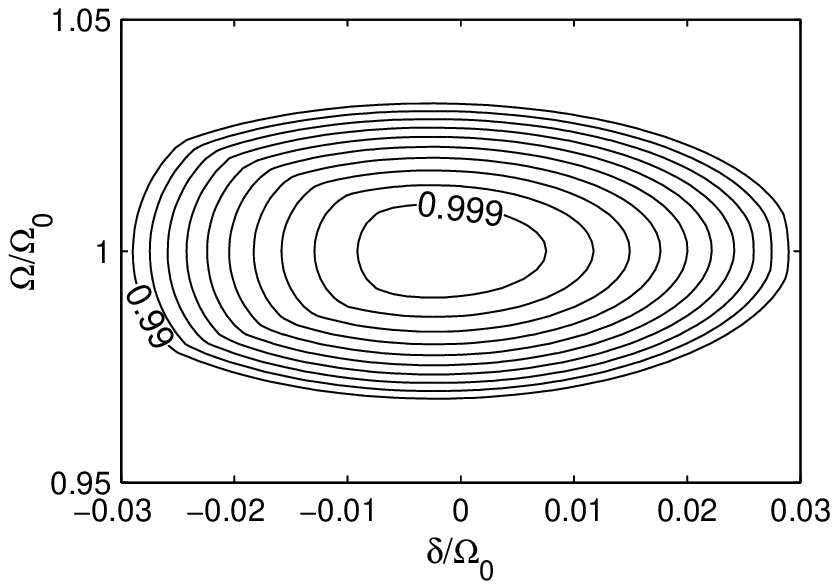}&%
    \includegraphics{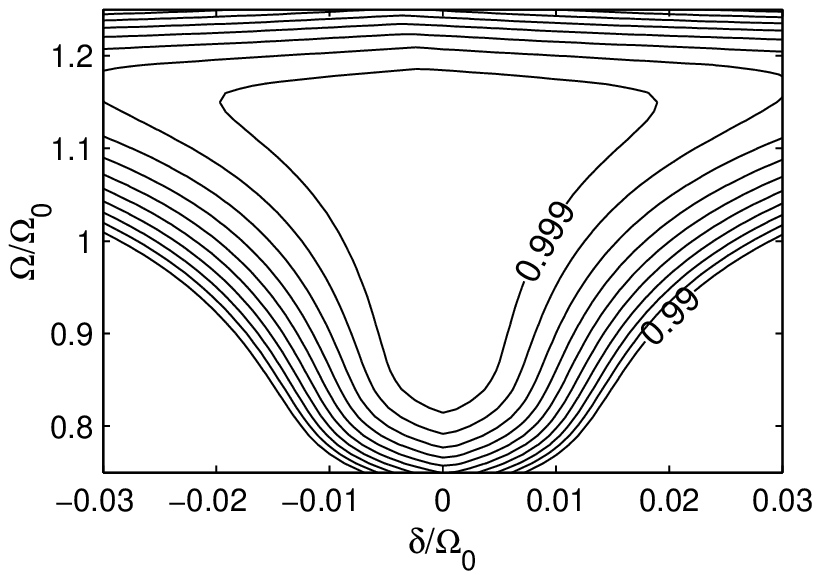}%
  \end{tabular}
  \caption{%
    Calculated worst case fidelities \eqref{eq:fid} of two
    implementations of the controlled phase shift:
    a) The simple implementation \eqref{eq:phasegate},
    and b) the $\pulse_\text{CPS}$ pulse sequence \eqref{eq:cphase}.
    The fidelity is plotted as a function of
    $\delta^{(1)}=\delta^{(2)}$ and 
    $\Omega^{(1)}_0=\Omega^{(2)}_0$, both relative to $\Omega_0$,
    and with $g_{12}=100\, \Omega_0$. 
    Note the difference between the $\Omega$-axis limits of the two plots.
    It is clear from the plots that $\pulse_\text{CPS}$ achieves
    a high fidelity over a much larger parameter space.
    In particular, $\pulse_\text{CPS}$ is much less sensitive to
    variations in $\Omega$, while the sensitivity to variations in
    $\delta$ does not seem to be significantly improved.  
  }
  \label{fig:phasegate}
\end{figure*}

Implementing all the pulses of $\pulse^{(i,j)}_\text{CPS}$ by composite
\emph{BB1} pulses \eqref{eq:ubb1}, we do indeed obtain a very robust
implementation of the controlled phase shift as illustrated in
Fig.~\ref{fig:phasegate}.
To asses the gate performance we have compared the effect of the gate
to the desired gate operation, $U_\text{CPS}$, in terms of the worst
case fidelity, defined as the minimal overlap between the actual
outcome of the pulse sequence and the desired outcome of the gate
operation:
\begin{equation}
  \label{eq:fid}
  \fid(U_\text{CPS},{\pp{}{}{\mu}})= \min_{\ket{\psi}\in \hilb} 
  \left|\Braket{\psi\left|
        U_\text{CPS}^\dag \, {\pp{}{}{\mu}}\right|\psi}\right|^2.
\end{equation}
Since we know that the starting point of the gate operation will be a
superposition of the ground hyperfine states, we have not minimized
the expression \eqref{eq:fid} over the full Hilbert space, but rather
restricted $\hilb$ to the qubit space.
Note that this modification ensures that any population in the excited
state after the gate operation is counted as a loss of fidelity as it
should be.
The computation of the fidelity is discussed in more detail in the
appendix.

As we see from Fig.~\ref{fig:phasegate} the pulse sequence
$\pulse_\text{CPS}$ obtains high fidelities over a much larger
parameter space than the simple gate operation described by
Eq.~\eqref{eq:phasegate}. 
This is highly desirable, as the minimal fidelity among the included
instances determine the scale-up needed to perform error correction
\cite{steane02:overh_noise_thres_fault_toler}.
Not too surprisingly, the sensitivity to variations in $\Omega$ is
improved the most, as this is the type of error best dealt with by the
BB1 pulse sequence.
For realistic parameters of the REQC system, a reduced sensitivity to
$\delta$ variation would be more useful; whether this can be achieved
by means of composite pulses is a point of further study.

\section{The bus architecture}
\label{sec:bus-architecture}
As the architecture of an REQC system can be chosen at will, the
question remains of which architecture to choose.

\begin{figure}[htbp]
  \centering
  \includegraphics{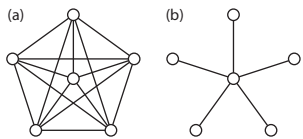}
  \caption{Two possible coupling topologies for REQC systems: a)
  cluster topology and b) star topology}
  \label{fig:starclust}
\end{figure}

The fully interconnected ``cluster'' architecture suggested in the
original REQC proposal of course has the minimal topological distance
between qubits.
On the other hand, a star topology with one central qubit coupled to
the $n-1$ remaining qubits, as illustrated in
Fig.~\ref{fig:starclust}, would reduce the number of required
couplings from $n (n-1)/2$ to $n-1$, thus increasing the number of
available instances in a given crystal, while still maintaining a
topological distance of only $2$.

Since the outer qubits in the star topology are not directly coupled,
two-qubit gates between those must be mediated by the central
qubit acting as a \emph{bus}.
To be specific, a bus-mediated controlled not gate can be constructed
as
\begin{equation}
  \includegraphics{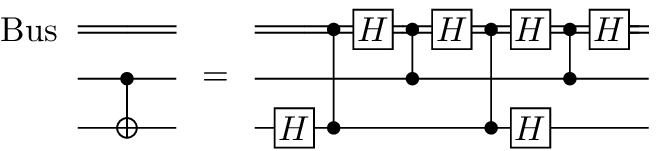},
\end{equation}
with the vertical lines signifying controlled phase shifts.

In addition to the better scaling properties of the bus architecture,
its main advantage is that the bus qubit is a participant of all
multi-qubit gates.
This fact can be used to ease or improve the implementation of such
gates: As an example, four times as many pulses are needed on channel
$j$ as on channel $i$ with the proposed implementation of the
controlled phase shift \eqref{eq:cphase}. If channel
$j$ is chosen as the bus channel, a dedicated laser system can
speed up the application of these pulses as compared to a tunable
laser system able to address any channel.

\section{Preparation and detection of maximally entangled states}
\label{sec:prep-detect-cat}
To demonstrate the viability of the REQC concept, and in particular the
bus architecture, we propose to perform an experimental preparation
and detection of a maximally entangled state.

We will use an REQC system with star topology: one central qubit
coupled to $n-1$ outer qubits.
Starting with all $n$ qubits in their $\ko$ state, we apply a
composite pulse Hadamard operation to the central qubit followed by
controlled not operations on all the outer qubits controlled by the
central qubit, thus transferring the system to the maximally entangled
state
\begin{equation}
  \label{eq:wineinit}
  \ket{\Psi_0}=\tfrac{1}{\sqrt{2}}\left(\ko^n+\kl^n \right),
\end{equation}
which corresponds to a superposition of the total pseudospin pointing
straight up and straight down.

The following algorithm for detecting a population of the cat state
$\ket{\Psi_0}$ is very similar to the method used by the group of D.
Wineland to detect a maximally entangled state of four ions in a
linear Paul trap \cite{sackett00:exper_entan_four_partic}:
By rotating the state $\ket{\Psi_0}$ through an angle $\phi$ around 
the $\vhat{z}$-axis we accumulate different phases on the pseudospin
components:
$\ket{\Psi_1}=\tfrac{1}{\sqrt{2}}\left(\ko^n+e^{-i \phi n} \kl^n
\right)$.
An additional rotation 
by $\pi/2$ around the $y$-axis now yields a state $\ket{\Psi_2}$ with
an expected parity, $P=\Pi_i (\sigma_z)_i$, given by
\begin{equation*}
  \label{eq:wineosc}
  \Braket{\Psi_2 | P | \Psi_2}=\cos(n \phi),
\end{equation*}
the detection of the $n \phi$ dependency thus signifying that the
maximally entangled state has been populated
\cite{bollinger96:qptim_frequen_measur_with_maxim}.

In a single-instance quantum computing system, such as the ion trap
setup used in Ref.~\cite{sackett00:exper_entan_four_partic}, we could
measure the expectation value of the parity as a statistical average
over many repetitions of the procedure described above: 
after each run we could simply measure the state of each
qubit, and subsequently compute the parity.
Since measurements in the REQC system yields an ensemble average, this
approach would not be applicable here: we cannot find the expectation
value of the parity from the ensemble averages of the single qubit
parities, $\expval{(\sigma_z)_i}$, which are $0$ as inspection shows.

Instead we let the bus qubit acquire the parity unitarily: by
sequentially applying controlled not operations from each outer qubit
to the central qubit we make the central qubit end up in the $\kl$
state in the case of odd parity and in the $\ko$ state in the case of
even parity.
After this, the ensemble average of the bus qubit population yields
the expectation value of the parity.

As this section illustrates, readout from an ensemble quantum
computer is conceptually somewhat more complicated than readout from a
single quantum computer.
It is worth noting, however, that unlike many other ensemble quantum
computing proposals, REQC instances all start in the same pure state:
if we successfully employ error correction during a computation all
instances will end up in the same pure state, allowing us to read
out the ensemble averages with high signal to noise ratio.
Perhaps surprisingly, the readout can almost always be performed by
tricks similar to those employed to detect the maximally entangled
state: Ensemble quantum computing is almost as powerful as general
quantum computing.  In particular, all problems which may be expressed
in terms of the hidden subgroup problem (such as Shor's factoring
algorithm) can be solved using an ensemble quantum computer
\cite{nielsen00:quant_comput_quant_infor}.

\section{Conclusions and Outlook}
In conclusion, we have shown that, in the absence of decay and
decoherence, it is possible to implement robust high-fidelity gates for
the REQC system.
Specifically, the phase compensated controlled phase gate based on
composite pulses \eqref{eq:cphase}, achieves worst case gate
fidelities above $0.999$, even with the coupling strength varying up
to $10\%$ between instances and channel widths of several percent of
the Rabi frequency of the field used to manipulate the system.
Furthermore, we have pointed out that using a bus based architecture
will simplify implementation by allowing the use of an asymmetric
laser setup.

The number of instances of a bus based REQC system scales as $p^n$
where $n$ is the number of qubits per instance and $p$ is the
probability of a random ion being coupled to a member of a given
channel.
In the regime currently being investigated experimentally, $p$ is
several orders of magnitude less than $1$.
The value of $p$ is affected by $\gt$ and channel width, which is why
we have to use robust gates rather than narrow channels and high
threshold coupling strengths. 
Higher values of $p$ could be obtained by increasing the ion density,
which would, however, cause a decrease in coherence times. 
By using structured doping techniques it might be possible to obtain a
higher effective $p$ without this adverse effect.
%
Another approach to obtaining higher effective $p$ would be 
to use multiple channels for each qubit by guaranteeing each instance
to have exactly one member ion from a group of channels
assigned to each qubit.

The instance identification protocol described in
Sec.~\ref{sec:reqc-proposal} could be made much more efficient: Since
the system starts in a pure state (all ions in the channels in their
$\ko$ state), and also ends in a pure state (all instance members in
their $\ko$ state, and all other ions from the initial channel
populations in their $\kaux$ state), the selection could theoretically
be performed unitarily.

\appendix*

\section{\label{sec:fidel-unit-oper}Fidelity of unitary operations}
We wish to compare unitary operators $U$ and $U_0$, by determining how
closely $U_0^\dag U$ resembles the identity on the Hilbert space
$\hilb$. This can be expressed in terms of the worst case fidelity:
\begin{equation}
  \label{eq:fidrepeat}
  \fid(U_0,U)= \min_{\ket{\psi}\in \hilb} 
  \left|\Braket{\psi\left|U_0^\dag \, U\right|\psi}\right|^2.
\end{equation}
The fidelity can be computed as follows: $U_0^\dag U$ is unitary and 
can consequently be formally diagonalized with eigenvalues $e^{i \phi_j}$,
$j=1,\ldots, n$ so that $0 \le \phi_1 \le \ldots \le \phi_n \le 2
\pi$.
Introducing the maximal eigenvalue phase distance 
$\Delta\phi_\text{max}=\max(\{\phi_{j}-\phi_{j-1}\}_{j=2,\ldots,n}
\cup \{2 \pi +\phi_1-\phi_n\} )$, 
the fidelity over $\hilb$ is given as
\begin{equation}
  \label{eq:fidresult}
  \fid(U_0,U)=
  \begin{cases}
    \cos^2(\Delta\phi_\text{max}/2) 
    &\text{ if $\Delta\phi_\text{max} \ge \pi$,}\\
    0 &\text{ otherwise.}
  \end{cases}
\end{equation}

To see this, we expand the state vector $\ket{\psi}$ on the eigenbasis
$\{\ket{j}\}$ of $U_0^\dag U$: $\ket{\psi}=\sum_j c_j \ket{j}$.
The fidelity then takes the form
\begin{equation}
  \label{eq:regne}
  \fid(U_0,U)
  =\min_{p_j} \left| \sum_j p_j\,e^{i \phi_j}\right|^2,
\end{equation}
with the minimum taken over all non-negative $p_j=|c_j|^2$, so that
$\sum_j p_j=1$.

Eq.~\eqref{eq:regne} allows us to interpret the fidelity geometrically
in the complex plane, as the set of points $\{\sum_j p_j \exp(i
\phi_j)\}$ form a convex polygon with vertices in the eigenvalues
$\{e^{i \phi_j}\}$ on the unit circle.
The fidelity corresponds to the square of the minimal distance from
$0$ to this polygon. If the polygon is constrained to one half-plane,
this will be $|e^{i \phi}+e^{i(
  \phi+\Delta\phi_\text{max})}|^2/4=\cos^2(\Delta\phi_\text{max}/2)$.
If the polygon is not restricted to one half-plane, it will cover the
origin, and the fidelity will be $0$.

Note that this method relies on the minimization being performed on
the whole Hilbert space. 
If this is not the case the method is not applicable, and in
Sec.~\ref{sec:full-fledged-gate} where the minimization is carried out
over a subspace of the full Hilbert space, we have resorted to a
numerical search.

\begin{acknowledgments}
We thank S. Kr{\"o}ll for stimulating discussions and comments on the
manuscript. 
This research was funded by project REQC of the 
IST-FET programme of the EC.
\end{acknowledgments}


\end{document}